Original article

# The effect of the Indonesian higher education evaluation system on conference proceedings publications

Philip J. Purnell [1,2]

[1] Centre for Science and Technology Studies,
Leiden University,
P.O. Box 905, 2300 AX Leiden, The Netherlands

[2] Knowledge E, Office 4401, X2 Tower, JLT, 488239 Dubai, United Arab Emirates

Corresponding author: Philip J. Purnell
Centre for Science and Technology Studies,
Leiden University,
P.O. Box 905, 2300 AX Leiden, The Netherlands
Tel: +971 50 552 9356
p.j.purnell@cwts.leidenuniv.nl
ORCID: 0000-0003-3146-2737

## Abstract

Indonesia has exhibited a unique pattern of conference proceedings publishing that was distinct from global and regional norms. Conference proceedings are now an integral part of the major citation databases, and this study examined their progressive coverage in the context of academic career advancement and Indonesian government policy. The results of analyses in Web of Science (WoS), Scopus and to a lesser extent, Dimensions showed an increase in the proportion of conference proceedings far in excess of global norms and not seen in any other South East Asian country. In the conference series containing most Indonesian proceedings papers, there was a recent increase in representation from Indonesia and an increase in the proportion of those conferences hosted in Indonesia. Local policy documents and guidelines from 2012 and 2014 appeared to encourage academics to increase research outputs in high impact internationally indexed sources but did not make any distinction between document types. The conclusion suggests that scholars might have found advantage in advancing through the hierarchy of academia through publishing conference papers that were quicker and easier to publish than journal articles. The study is important to policy makers in the area of research evaluation because it highlights potential changes in academic publishing behaviour by those being assessed.

# Introduction

This study was originally intended to be an analysis of the proportion of conference proceedings papers among scholarly publications. It would be expected to find varying rates of conference papers according to subject field, publication year and geographical region. As the relationship between conference papers and subject fields have been extensively studied, the focus of this paper was to determine whether the location of author played a role and how that developed over time.

Whilst gathering data for the study, I noticed unusual patterns in the countries of the southeast Asian Nations (ASEAN) and therefore decided to focus on this group of ten countries as a region. The ten component countries display rather different patterns, the most interesting being Indonesia because of an evident wholesale shift towards publishing conference proceedings at the expense of journal articles. I tested several potential explanations for this trend and concluded the most likely being a response to a change in the national system of evaluating and promoting individual academics.

A new set of government publishing guidelines has been recently released in Indonesia related to the promotion of scholars through the academic hierarchy that gives slightly higher weighting to journal articles than conference proceedings. Earlier guidelines from 2012, 2014 and 2017 did not make that distinction and it is therefore interesting to investigate whether the earlier policies could have directly led to a preference in Indonesia for publishing conference papers as these are more easily accepted and pass more quickly through peer review and into publication.

# Literature review

An evolutionary model of scholarly communication was described several decades ago in which academic research is communicated first as personal correspondence and in subsequent incremental forms that include open letters, conference proceedings and finally the journal article (Garvey et al. 1972a). A detailed study on the publication behaviour of 12,000 scientists over a five-year period demonstrated that the majority of conference material presented at national meetings in America was later published as a journal article (Garvey et al. 1972b) and this pattern became broadly accepted as common practice in the exchange of scientific information between researchers (Drott 1995). Later evidence challenged this view with conference proceedings seen by some as an end product and being accepted as evidence of scholarly activity by university tenure and promotion committees (Drott 1995), rather than being merely a step on the way to a journal article.

The advent of the Internet spawned fundamental changes to communication of academic research findings with some results appearing on the web before they were published in journals or as conference proceedings (Goodrum et al. 2001). Work comparing conference papers with journal articles showed that conference papers are generally shorter (Gonzalez-Albo & Bordons, 2011), less cited (Drott 1995) than journal articles and their citation peak is shorter lived (Lisée, Larivière & Archambault, 2008). In many fields assessment for promotion gives higher weighting to journal articles, which serves as an incentive for scholars to adapt conference proceedings for subsequent journal submission. However, in other fields proceedings papers and journal articles are seen as different expressions of the same work (Bar-Ilan 2010) revealing a variance in publication behaviour between fields.

There is broad variance in the relative importance of proceedings papers depending on the field of study. One study showed that roughly half the papers in ISI Proceedings – the first conference proceedings database and later known as the Conference Proceedings Citation Index (CPCI) – were assigned to the field of Engineering and that this share increased from 43% in 1994 to 61% by 2002 (Glänzel et al. 2006), more than six times the corresponding share found in the Web of Science (WoS) journals. In the same study, the proportion of proceedings papers classified in the field of physics grew from 25% to 32% over the eight-year period while the corresponding proportion of journal articles in the WoS remained stable at around 13%. Similarly, the engineering field was singled out as one in which proceedings papers receive a higher proportion of citations indicating that conference material is of greater import than in other fields (Lisée et al. 2008). This was especially the case for computer science papers.

Computer science has also been identified as a field in which conference proceedings are a major venue (Bar-Ilan 2010) for disseminating research findings. Bar-Ilan (2010) pointed out that conference papers from Lecture Notes in Computer Science (LNCS) were already indexed in the Science Citation Index before the addition of the CPCI to the WoS in 2008. In a study comprising interviews with authors and journal editors in the software engineering field, conference papers were thought to be shorter than journal articles and contain only the exciting part of the study intended to keep a specialized audience abreast of novelties in their field (Montesi & Owen 2008). Journal articles were found to be a more mature product designed to enable readers to replicate results and form part of an archive. Furthermore, conference papers were often reworked and later published as journal articles. An extensive study of citations to non-WoS literature in 3 social science fields showed that only 2% of cited references to academic literature outside the WoS were to conference papers in psychology, only 1% in political science and even less in economics (Nederhof et al. 2010) showing very limited influence of non-WoS literature on highly cited works.

## Data sources

A Dutch study concluded in 2007 that it was feasible to expand the WoS to include additional conference papers to provide better coverage in the field of computer science provided some technical issues were addressed such as the treatment of different versions of the same study (Moed & Visser 2007). In 2008, Thomson Reuters merged the content from its two conference products, the Conference Proceedings Citation Index – Science (CPCI-S) and the Conference Proceedings Citation Index – Social Sciences & Humanities (CPCI-SSH), formerly known as ISI Proceedings Science and Technology (STP) edition and ISI Proceedings Social Science and Humanities (SSHP) edition into the WoS. Until then, the WoS had comprised 3 journal indices; the Science Citation Index Expanded (SCIE), Social Sciences Citation Index (SSCI) and the Arts & Humanities Citation Index (A&HCI) each consisting of metadata from academic peer reviewed journals. Although a limited number of conference proceedings were already published in journals indexed in the database, it was the incorporation of the CPCI that contributed the majority of the conference literature to the WoS.

It was this addition of conference proceedings into the Web of Science in 2008 that provided an opportunity to study these as a unique source of scholarly material.

At about the same time, journal articles that had been adapted from meeting presentations had their document types changed from 'paper' to 'proceedings paper', the same name given to the majority of the publications in the CPCI. All documents indexed from conference material were thus assigned the document type 'proceedings paper' although they may have originated either as journal articles that made explicit reference to initially having been presented at a scientific meeting, or may instead have been published in a 'book' of proceedings. Proceedings papers published in journals have been found to vary in their proportion with respect to regular articles by field with highest proportions found in computer science/information technology and applied physics (Zhang & Glanzel 2012). Conference proceedings in journals can be further categorized into those accepted in ordinary issues and those published as special editions or monographs. Those published in monographs have been found in library and information science to be shorter, have fewer references, pass more quickly through peer review and receive fewer citations (González-Albo & Bordons 2011).

These nuances in publication behaviour, changes in documentation types and evolution of the WoS coverage over time mean that any study of publications is susceptible to an element of misinterpretation of the results owing to the idiosyncrasies of the database. For this reason, it can be useful to replicate part of the study using other databases such as Elsevier Scopus released as a more inclusive citation index in 2004, or Dimensions launched in 2018 by Digital Science. The increase in published conference proceedings from Indonesian authors observed in the WoS was so surprising that I aimed to check whether WoS expansion might be partially responsible. The WoS CPCI has grown modestly in the past 10 years but in order to rule out effects of the WoS coverage expansion accidentally indexing a large number of Indonesian conference papers, I used Scopus and Dimensions to try to replicate the results.

Elsevier launched its proprietary citation index, Scopus, in 2004 which now covers more journals than Web of Science (Mongeon & Paul-Hus 2016). However, the overlapping coverage of conference proceedings between WoS and Scopus is limited. Indeed, WoS appears to cover many conference proceedings that are not indexed in Scopus, and Scopus covers many proceedings not covered by WoS (Visser et al. 2019). Dimensions, a relative newcomer to citation indexing, offers a free version including access to just over 100 million scholarly records. Users can identify more than 5 million conference proceedings among the records without requiring an organizational subscription. Dimensions coverage is comparable with Scopus, indeed at least 90% of Scopus indexed papers were found in Dimensions with the exception of the most recent year in which coverage dropped to about two-thirds (Thelwall 2018) probably due to a longer time lag in indexing in Dimensions. The overlap of documents with Crossref indicates that Dimensions relies heavily on Crossref for its data but many Scopus indexed conference proceedings are absent in Dimensions (Visser et al. 2019). This means the coverage of the two databases is independent and both should be used to verify the results observed in the WoS.

Even if all the bibliometric sources produced the same result and showed rapid increase in conference publishing by Indonesian authors, it would still be affected by the rate at which all the databases added proceedings to their listings. Any entity publishing heavily in those newly added proceedings would appear to have an increased output regardless of whether their publications had actually increased. Therefore, simply reporting the number of conference papers was not sufficient to detect a real change in publishing behaviour by Indonesian authors. Instead, it was necessary to calculate the proportion of publications that were conference papers. Comparing this proportion between countries and how the relative proportions changed over time then revealed how the authors' behaviour differed between countries.

The remainder of this analysis focuses on the conference proceedings published by scientists in Indonesia.

**Policy context**

This paper analyses the conference proceedings output from the emerging economies in Asia, with a special focus on Indonesia, in an attempt to determine whether there is a causal relationship between national science policy and publication behaviour. In many systems, scholarly publication has become the currency that academics are expected to produce with incentives for increasing output. It stands to reason therefore that people will choose the path most likely to produce a desirable result and if one type of publication is easier or quicker than another it will become an attractive route. In some fields it is easier to have a conference paper accepted and one might therefore expect to see an increase in proceedings papers in systems that reward raw publication output. However that is not necessarily a universal relationship given that in computer science publishing a proceedings paper may be harder than publishing a journal article (Ulusoy 1995).

I obtained a number of local policy documents from Indonesia published between 2010 and 2019 and found that they increasingly encouraged scientists to publish papers in internationally indexed journals. (Ministry of Education and Culture 2014; Ministry of Higher Education 2012; Ministry of Research & Higher Education 2019; Ministry of Research Technology and Higher Education 2017). The documents also provided means for scientists to apply for funds to visit overseas conferences in recognition of the difficulties and resources required for international trips. It is reasonable to suppose that it is easier for local scholars to attend conferences hosted in the country they live in, and to present their findings there. This study looks at the author affiliations and the locations of conferences to examine any obvious connection.

Researchers have studied the link between science policy and publication behaviour in other parts of the world and described such relationships. For example, research output and citation impact in Denmark reversed a downward trend seen in the 1980s and showed sustained improvement over the subsequent two decades with respect to other developed nations. The precise relationship was too complicated to establish but the study showed that the timing of the upturn coincided with a major change in the way public funds were used for R&D as part of the Danish Globalisation Strategy (Aagaard & Schneider 2016). Another example is the United Kingdom's periodic national research assessment exercise (RAE) and later research excellence framework (REF) which have been linked to a reduction in published books, book chapters, conference papers, and reports, indeed all types of publication except the journal article (Marques et al. 2017). These authors described an increase in the proportion of journal articles submitted by the universities for assessment from 57% in the 2001 RAE to 77% in the 2014 REF as a 'self-fulfilling prophesy', since institutions have selectively submitted document types likely to return a high impact score.

As a third example, the Australian government introduced a productivity component into its funding distribution formula during the 1990s. Researchers and their institutions received financial reward every time they published an article in an international peer reviewed journal indexed in the WoS, effectively putting a monetary value on a publication (Butler 2003a). Butler claimed that the policy stimulated a behavioural change in Australia's scientists (Butler 2003b) whose output increased in WoS journals following the introduction of this policy, but that the increase occurred chiefly in the lower impact journals in which it might have been easier to have a paper accepted and thereby lowering their average citation impact. Subsequent studies conducted with the benefit of additional data and article-level impact indicators suggested Butler's conclusions were premature because the time frame for her study was too early in relation to policy changes. (van den Besselaar et al. 2017). Butler responded to this criticism saying that her conclusions were not premature because Australian universities were aware of, and already responding to changes in policy well before those policies were formally implemented (Butler 2017). She also rejected the link between low impact journals and low quality research papers implied by van den Besselaar and colleagues. Butler pointed out that Australian academics increased their publication rates in high impact journals during the period studied, but not as quickly as they increased their submissions to lower impact journals.

**Regional focus**

Publishing behaviour is subject to universal forces such as the requirement on academics to publish their work and demonstrate some form of impact. Regional differences in publication behaviour can provide an interesting basis for more specific study, for example the fast-growing emerging regions such as those of South East Asia where 10 countries, Brunei Darussalam, Cambodia, Indonesia, Laos, Malaysia, Myanmar, Philippines, Singapore, Thailand and Vietnam, have formed the Association of Southeast Asian Nations (ASEAN). The organization was founded in 1967 (ASEAN 2019) in order to promote peace, collaboration, regional research, and common social, cultural and economic values. As the preliminary findings suggested that conference proceedings research output was rising fast in the region, I chose the ASEAN countries for deeper analysis in this study.

In 2011, Nguyen and Pham studied the scientific output and impact of the ASEAN countries and described four distinct groups with Singapore alone in the most advanced bloc, Malaysia and Thailand in the second, Indonesia, The Philippines and Vietnam in the third, and finally Brunei, Cambodia, Laos and Myanmar making up the fourth cluster (Nguyen & Pham 2011). These researchers also found correlations between scientific output and both the Knowledge Index and the Knowledge Economy Index published by the World Bank in 2008 (Chen & Dahlman 2005) each of which confirmed the four distinct groups of ASEAN countries observed by Nguyen and Pham. The rate of economic growth of countries such as Singapore has been linked to the investment in research and development (R&D) (Nguyen & Pham 2011). A large 40-country study of investment and publication data demonstrated a positive correlation between R&D spending and research publication output (Meo et al. 2013).

As part of a UNESCO report, Moed and Galevi described a bibliometric model to group 25 Asian countries' relative stage of development by organizing them into three clusters based on the proportion of their internationally co-authored papers and the geographical location of the collaborating countries (Moed & Halevi 2014). The first group includes only Singapore among the ASEAN countries and whose papers were often co-authored with researchers in China, Hong Kong and Macau. The second cluster also only features one ASEAN country, Malaysia, which is more linked to research groups in India, Pakistan and Iran, while the third group contains the remaining members in a South East Asian cluster. These authors also draw a link between the state of a country's development and the ratio of the number of its doctoral students to the number of its publications. For example, Japan, a developed country has roughly the same number of doctoral students as Indonesia but produces 100 times the number of publications as Indonesia (Moed & Halevi 2014).

**Methods**

For the WoS studies, I used an in-house version of the Web of Science database hosted by CWTS at Leiden University that comprises five citation indices; the Science Citation Index Expanded (SCIE), Social Sciences Citation Index (SSCI), Arts & Humanities Citation Index (A&HCI), Conference Proceedings Citation Index – Science (CPCI-S) and the Conference Proceedings Citation Index – Social Sciences & Humanities (CPCI-SSH) and is collectively referred to as WoS 5-ed meaning five editions. Neither the Book Citation Index (BkCI) nor the Emerging Sources Citation Index (ESCI) were included in the study because they are not hosted in the system. Likewise, I used an in-house version of Scopus extracted in May 2018 and Dimensions extracted in December 2018 in order to reduce the risk of conclusions being drawn due to any artefact in the WoS database.

I counted the conference papers in WoS from CPCI whereas from Scopus and Dimensions the document types 'proceedings paper' and 'proceeding' were used respectively. In each case, I calculated the proportion of conference papers by dividing the conference papers by the total output.

I calculated the rate of growth in publication output using compound annual growth rate (CAGR) as:

$$CAGR = \left(P_e/P_s\right)^{(Y_e - Y_s)} - 1$$

where $P_e$ is the number of papers in the end year of the analysis and $P_s$ the number of papers in the start year, and $Y_e$ is the end year and $Y_s$ the start year of the period for analysis.

To determine Indonesia's rank in conference output, I ranked the country affiliations on the proceedings papers and noted Indonesia's position. I identified the location of the conferences from the publisher websites, e.g. AIP Publishing LLC (AIP 2019; AtlantisPress 2019; IOPPublishing 2019) and listed the proportion of conferences that were hosted in Indonesia. For detail on Indonesian scientific publishing policy, I translated original government reports and guidelines using Google Translate and then checked my/our understanding with the help of a native speaker who is also a subject matter expert.

# Results and Discussion

## Global growth trends

The number of conference proceedings records in WoS was fairly stable for the initial period and grew 38% in the first eleven years (Fig. 1) which was in line with the rest of WoS as demonstrated by the stable proportion of CPCI records in the WoS 5-ed.

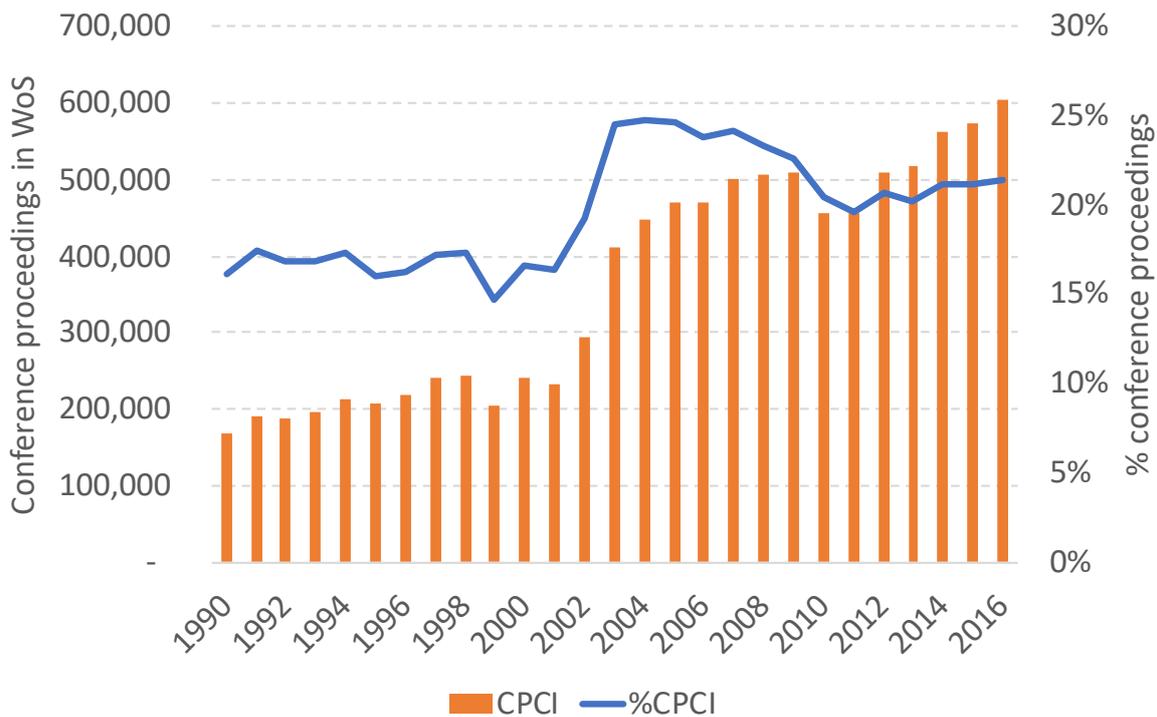

Figure 1.   Number and proportion of CPCI papers in WoS 5-ed

Thereafter followed a steep increase in both records (77%) and proportion between 2001 and 2003, which showed that CPCI's contribution to the WoS rose from 16% to 25% in two years. In the subsequent fourteen years the number of CPCI records in WoS has grown only 48% and its proportion of the database reduced and stabilized at around 21%.

The relative proportions of the different document types have changed over time in the WoS. The number of conference papers indexed in the WoS showed steady growth for the initial ten-year period, that was followed by a steeper increase over the following five years. This steep increase in conference papers coincided with the launch of a CD Rom version of 'ISI Proceedings' in the early 2000s. The database owner, then Thomson Reuters, sought to create more value in this product by adding proceedings papers from conferences held in the preceding five years[1] in addition to those that were indexed annually. The additional conferences swelled the number of proceedings papers that resulted in a peak of 25% in the proportion of conference papers in the WoS in 2003. Growth tempered after 2009 as it was around this time that emphasis shifted towards the indexing of books and book chapters in preparation for the launch of the Book Citation index, which was in 2011, after which work resumed its usual pace on proceedings. In the most recent five years, the conference papers have reached a stable ratio of around 1:4 compared with journal papers as both grew at a similar rate.

---

[1] The author is a former employee of Thomson Reuters.

**Regional trends**

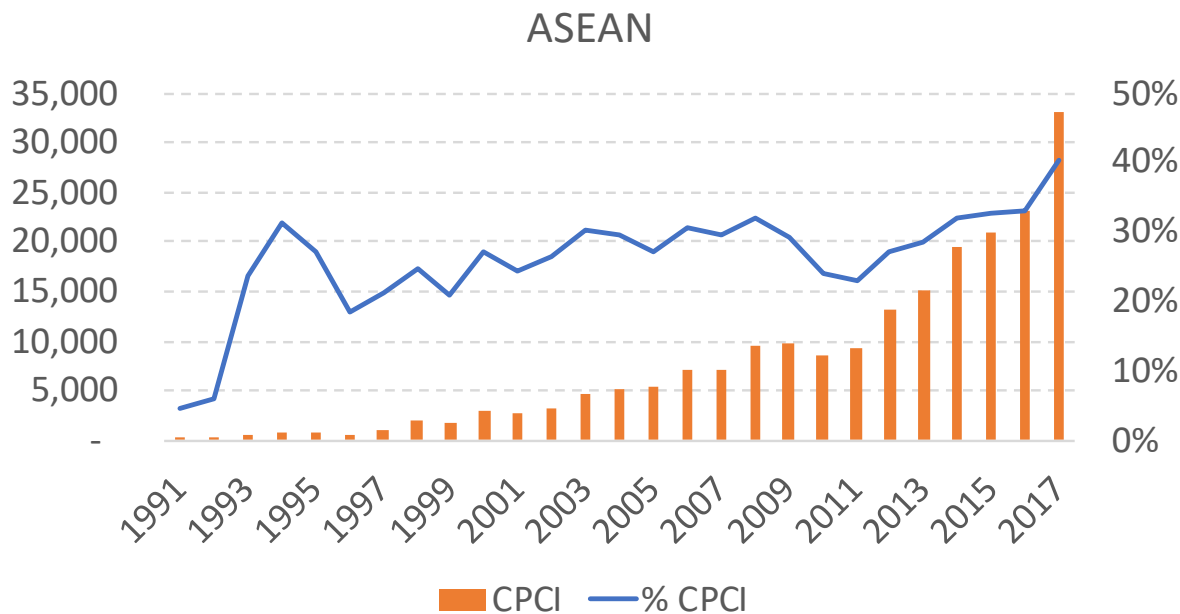

Figure 2.      Number and proportion of CPCI papers in WoS 5-ed in the ASEAN region

The growth in conference papers by ASEAN region authors was more sporadic (Fig. 2), starting from a very low base in 1991 and only surpassing 1,000 papers in 1998. The sudden growth between 2001 and 2003 seen in the global graph was obscured in the ASEAN regional data because of the rapid growth between 1997 and 1998, and between 1999 and 2000. The proportion of conference papers did grow from 24% to 30% between 2001 and 2003 coinciding with the increased proportion observed in the global figures. The dip in output and percentage conference papers between 2009 and 2011 was however present at both global and ASEAN levels.

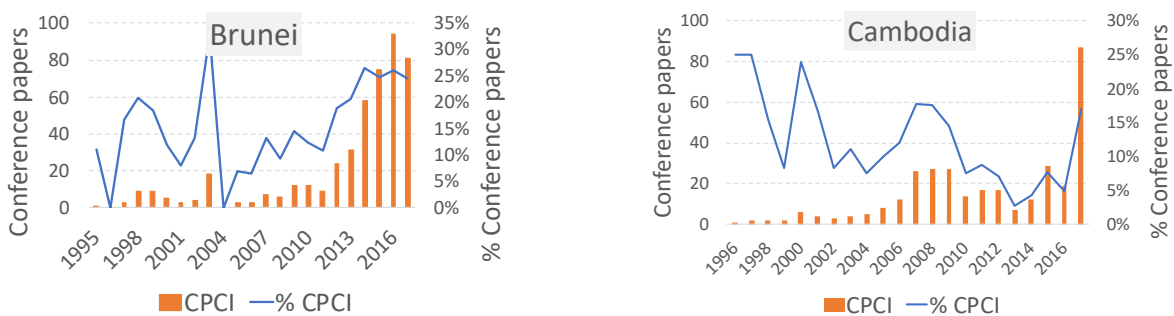

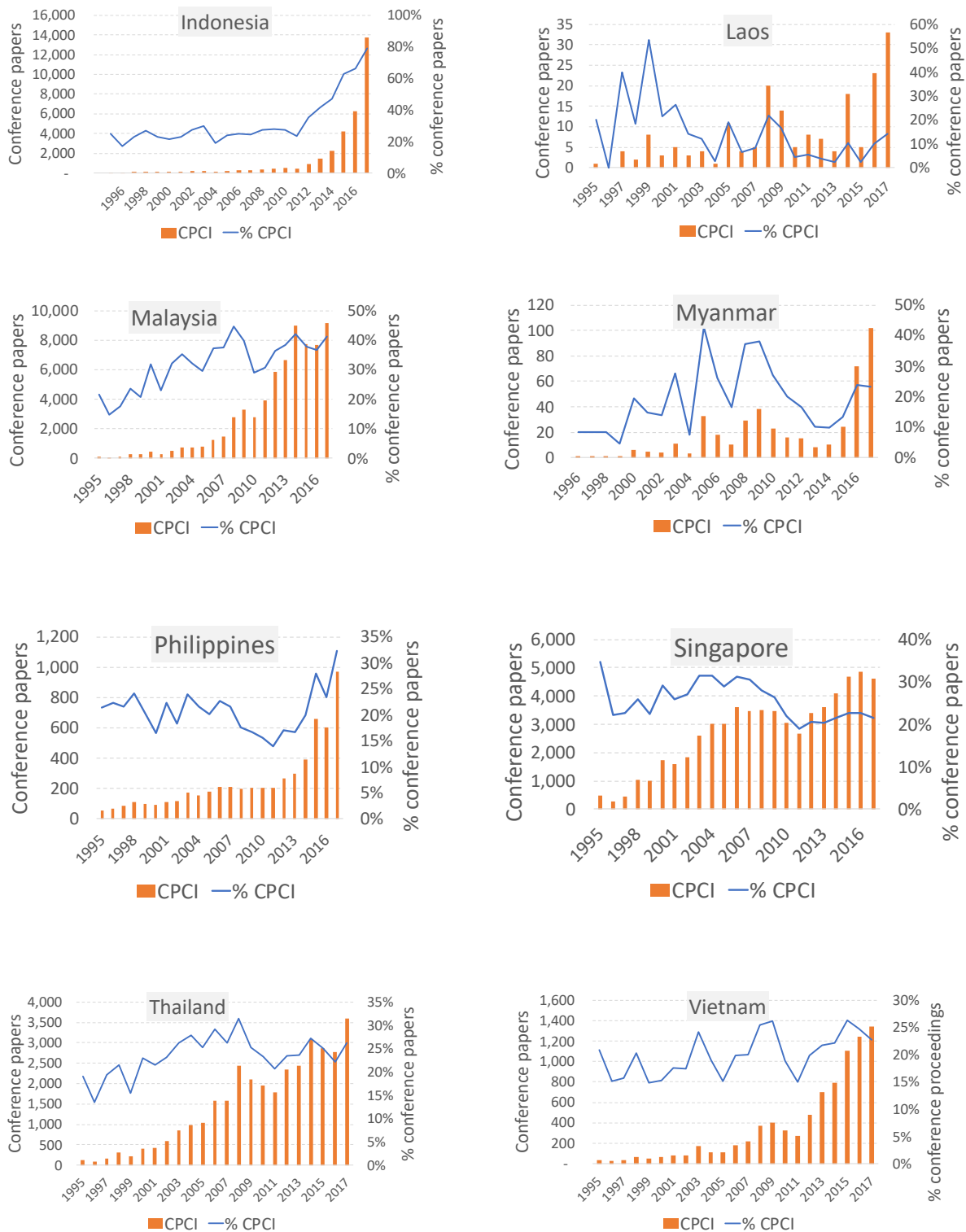

Figure 3. ASEAN countries' conference papers in WoS 2008-2017

The scientific output from different countries varies widely and the number of conference papers published by authors in the ASEAN region and their proportion of overall publications has been diverse with each country displaying its own unique pattern of growth. The corresponding graphs are shown in Figure 3 for each of the ten ASEAN countries showing the number of conference papers published and the proportion of that country's overall WoS publications that were conference papers.

The ten-year compound annual growth rate (CAGR) in conference paper output is presented for the ten ASEAN nations and compared with the ASEAN and world aggregate figures between 2008-2017 in Table 1. The individual counts for each year and country/region are displayed in Table A1 of the Appendix.

Table 1.    CPCI growth rates by country/territory

| Country/region | CAGR Conference papers 2008-17 (%) |
|---|---|
| Indonesia | 43.7 |
| Brunei | 29.7 |
| Philippines | 17.3 |
| Vietnam | 13.7 |
| Myanmar | 13.4 |
| **ASEAN** | **13.2** |
| Malaysia | 12.6 |
| Cambodia | 12.4 |
| Laos | 5.1 |
| Thailand | 3.9 |
| Singapore | 2.9 |
| **World** | **2.2** |

The ASEAN countries with highest CAGR in conference paper publications over the past ten years are Indonesia (43.7% growth per year) and Brunei (29.7%). Of these, Indonesia has both a sizeable body of conference papers for analysis, and a growth rate that was a clear outlier in its region and sufficiently interesting to study in further detail.

The Indonesian CAGR was more than three times the ASEAN average (13.2%) over the ten-year study period and almost twenty times the world average growth rate (2.2%). In 2017, almost four-fifths (79.2%) of indexed Indonesian papers in the WoS were conference proceedings – almost twice the corresponding proportion in Malaysia, which had the second highest proportion (41.3%), and the combined ASEAN countries (40.4%) and more than three times the world figure (22.0%).

Table 2. The ASEAN region with its populations and economies

| Country | Population (millions) 2017 | % ASEAN | % World | Country | GDP (billions) 2017 | % ASEAN | % World |
|---|---|---|---|---|---|---|---|
| Brunei | 0.4 | 0.1 | 0.01 | Brunei | 12.1 | 0.4 | 0.01 |
| Cambodia | 16.0 | 2.5 | 0.21 | Cambodia | 22.2 | 0.8 | 0.03 |
| Indonesia | 264.0 | 40.8 | 3.51 | Indonesia | 1,015.4 | 36.7 | 1.25 |
| Laos | 6.9 | 1.1 | 0.09 | Laos | 16.9 | 0.6 | 0.02 |
| Malaysia | 31.6 | 4.9 | 0.42 | Malaysia | 314.7 | 11.4 | 0.39 |
| Myanmar | 53.4 | 8.2 | 0.71 | Myanmar | 67.1 | 2.4 | 0.08 |
| Philippines | 104.9 | 16.2 | 1.39 | Philippines | 313.6 | 11.3 | 0.39 |
| Singapore | 5.6 | 0.9 | 0.07 | Singapore | 323.9 | 11.7 | 0.40 |
| Thailand | 69.0 | 10.7 | 0.92 | Thailand | 455.3 | 16.5 | 0.56 |
| Vietnam | 95.5 | 14.8 | 1.27 | Vietnam | 223.8 | 8.1 | 0.28 |
| ASEAN | 647.4 | 100.0 | 8.60 | ASEAN | 2,764.9 | 100.0 | 3.42 |
| World | 7,529.7 | | 100.0 | World | 80,934.8 | | 100.0 |

**Source: World Bank 2017**

The ASEAN countries comprise almost 650 million people (Table 2) making up 8.6% of the world's population and the combined economy of the ten members contributed 2.7 trillion dollars, equivalent to 3.4% of the global GDP in 2017 (World Bank Group 2019). The distribution of people and wealth within the region is uneven with the most populous country (Indonesia) home to 660 times the most sparsely populated (Brunei) and the Indonesian economy, which ranks 16[th] largest in the world, more than 80 times the size of the smallest (Brunei).

All of the ten countries that make up the ASEAN region have shown a steep increase in the number of conference papers their scientists have published during the study period. However, as we see from both the regional and world graphs this is not a phenomenon specific to any country or region. Rather it reflects a general increase in research output and coincides with the launch of the CPCI and consequent indexing of new conferences. The overall scientific output of nations has been studied elsewhere and is not the focus of this paper, which concentrates instead on the proportion conference papers among scientific publications.

Those countries showing sustained growth in the proportion of conference papers in their scholarly output over recent years were Brunei, Indonesia, Malaysia and Philippines. Other countries have demonstrated the opposite trend (Cambodia, Myanmar, and Singapore) with a decrease in conference material as a proportion of total production. The remaining countries (Laos, Thailand and Vietnam) have shown an inconsistent pattern of conference publishing. The growth rate in Indonesian conference papers far outstripped all other countries or regions in this study and has a large enough body of papers to pique interest in finding an explanation for the trend.

Other regional peers with sizeable conference outputs, namely Malaysia, Singapore and Thailand, all showed increases in raw conference proceedings output, and in Malaysia this was at the expense of other document types, but no country showed such an extreme increase as Indonesia. The pattern observed in Indonesia is a sustained and steep increase in published conference papers at the expense of other document types including journal articles and is distinct from the global picture and not mirrored by any of its ASEAN peers.

There are multiple potential explanations for the observed increase in conference paper output from Indonesia. Behavioural changes in academic publishing as a response to policy incentives are a potential explanation. However, first it was necessary to consider whether other explanations could be found, in particular the expansion of the database that was used for the study. The WoS comprised almost exclusively journal articles until the CPCI was incorporated in 2008. This merger added large numbers of conference proceedings and meetings abstracts to the database going back five years to 2003 thus enriching it both for discovery purposes and providing the means to better study publishing behaviour in those fields in which conference papers were more common.

The launch and subsequent rapid expansion of CPCI by indexing proceedings from additional conferences should naturally contribute to augmented counts and one would predict a somewhat turbulent expansion. It would be too difficult to coordinate indexing of new conferences in order to ensure an even geographical distribution of author affiliations. Therefore, some variance is to be expected, though that does not explain the unique pattern seen in Indonesia.

**Scopus and Dimensions**

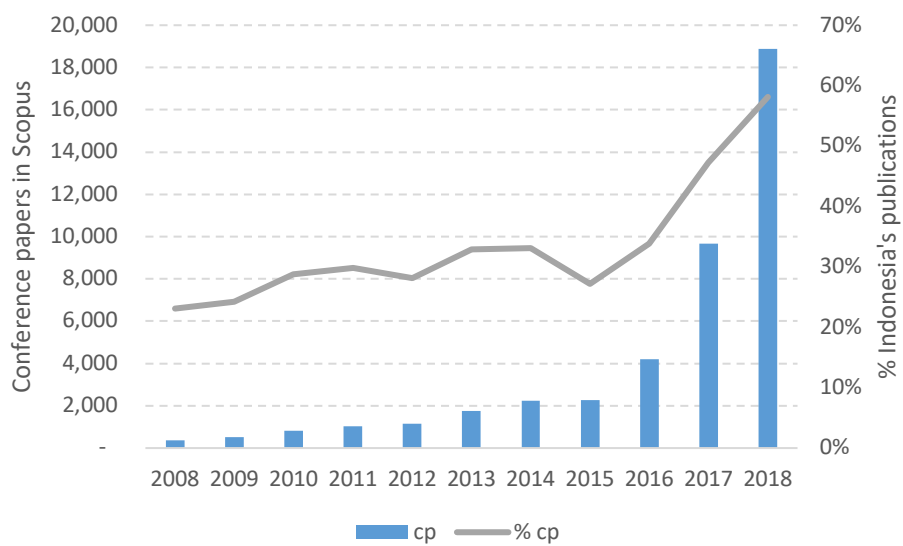

Figure 4. Indonesian conference papers in Scopus 2008-2018

The Scopus conference paper output from Indonesia increased more than six-fold between 2008 and 2015 (Fig. 4) but remained within a range of 23% - 33% of Indonesia's output in the overall database. In the following three years, the number of Scopus-indexed conference papers from Indonesia multiplied by more than eight times, increasing faster than the journal article rate. By 2018 Indonesia's conference papers accounted for almost three-fifths (58%) of the country's Scopus output.

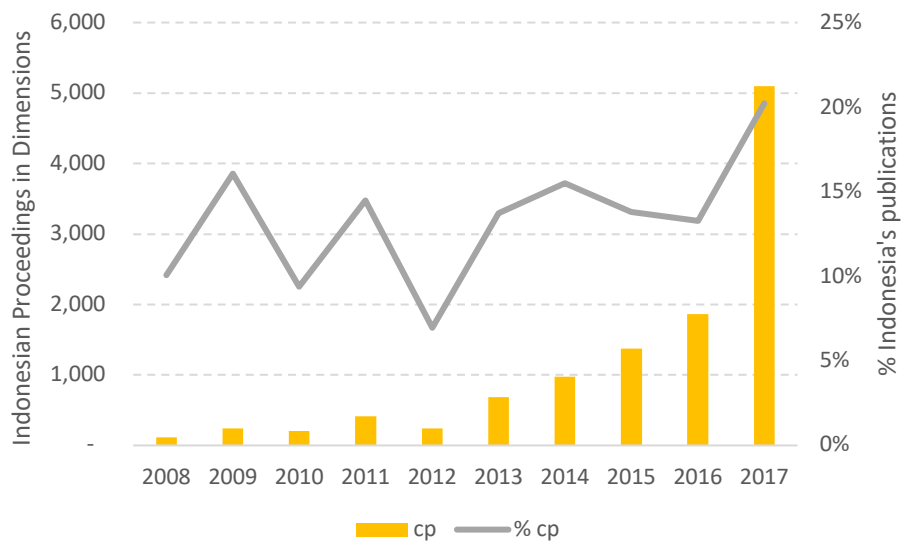

Figure 5.   Indonesian conference papers in Dimensions 2008-2017

The data in the Dimensions database (Fig. 5) showed the number of conference proceedings written by Indonesian authors going up and down until 2013 and oscillating between 7% and 16% of national output. By the most recent full year 2017, the share jumped to 20% accompanying a nearly four-fold increase in publications compared with 2015.

Although the pattern partially mimics that seen in Scopus in terms of growth, the increase is less pronounced in share reaching a maximum of only 20% in the most recent year. The reason for the lower proportion of Indonesia's conference papers in Dimensions might be due to a generally lower coverage of conference literature in the whole Dimensions database. The number of conference papers in Dimensions is lower than in Scopus and Web of Science, whereas the number of journal articles is greater.

In the Scopus database, the rate of change in Indonesia's conference papers showed a similar pattern to that seen in the WoS albeit less extreme and shifted approximately four years later. The relative delay of the phenomenon is possibly due to the later addition of conference papers to Scopus as part of the Scopus Conference Expansion Programme (Elsevier 2015) in which over 400,000 conference papers were added to the database during the years 2011-2014. This was several years after the CPCI was added to the WoS and would explain the time difference between the two databases in the surge of conference papers. The proportion of Indonesian conference papers seen in Scopus is not as remarkable as that observed in the WoS, but it is still high and has nearly doubled in the past three years, thereby confirming the existence of behaviour worthy of study.

Results from the Dimensions data mirrored those seen in the Scopus database and growth was especially prominent in the latest year although the proportion of conference proceedings remained far lower than in either Scopus or WoS reaching a maximum of only 20%. The lower proportion of conference papers in the Dimensions data might be due to greater coverage of journal articles globally combined with a smaller coverage of conference proceedings papers. We also have one extra year's data for Scopus – 2018 that is not yet available for Dimensions. In that year, there was a jump in Scopus from 48% to 58% conference papers and so it may be that we see a similar jump in the Dimensions data in due course.

The differences shown in Indonesia's conference publishing pattern between the three databases might be explained by differences in the coverage of each database and the precise timing of when conference series were indexed but fails to link publishing behaviour to artefacts in the data. This suggests that the most likely explanation for the observed increase in conference paper output from Indonesia can be found within the Indonesian academic community.

**Conference coverage expansion or organic Indonesian growth?**

One plausible explanation for the increase in Indonesian conference papers in the WoS is the natural expansion in CPCI (part of WoS) coverage of certain conference series that contain large numbers of Indonesian conference papers. A conference series is a grouping of published proceedings from conferences in a similar field. Many conferences series are indexed in the WoS and there are many that are not. The CPCI grows over time through routine addition of conferences and conference series to the database based on an evaluation process similar to that used to add journals to the WoS. Therefore, if a conference series not indexed in WoS and rich in papers from a particular country, were added to the WoS as a result of coverage expansion, then we would see an increase in output from that country. That increase would have been the result of the WoS coverage expansion, and not due to increased publishing activity from that country.

It is worth considering therefore, that Indonesia's increased representation in the CPCI might be due simply to the fact that the WoS has added a number of conference series to its database that contained large numbers of Indonesian papers prior to incorporation to the WoS. In order to test this explanation, I identified the six conference series with most Indonesian papers in the WoS in the last five years (Table 3), the period during which the major growth occurred. The aim was to determine whether these series were recently added to the WoS and if they contained large numbers of Indonesian papers prior to their indexation in the WoS. The results showed that each of the six series was already indexed in the WoS before any significant contribution from Indonesia, thereby dismissing the sudden addition of specific conferences to the database as a plausible explanation for the growth.

Table 3. Conferences by number and world share of proceedings papers from Indonesia (2013 and 2017).

| Conference | Proceedings papers 2013-17 | |
|---|---|---|
| | Indonesia | World share (%) |
| AIP Conference proceedings | 3,915 | 10.1 |
| Advances in Social Science Education and Humanities Research | 2,889 | 15.4 |
| Journal of Physics Conference Series | 1,599 | 5.5 |
| IOP Conference Series-Materials Science and Engineering | 1,401 | 9.6 |
| IOP Conference Series-Earth and Environmental Science | 972 | 15.3 |
| AEBMR-Advances in Economics Business and Management Research | 752 | 18.5 |

Table 3 shows six conference 'series' that each comprise proceedings papers from multiple conferences and are indexed in the CPCI. The series are those from which most Indonesian CPCI papers are registered show that in each case Indonesia based scientists account for between 5.5% and 18.5% of the total number of conference papers listed in each of the series. This resulted in between 752 and 3,915 papers indexed in the CPCI.

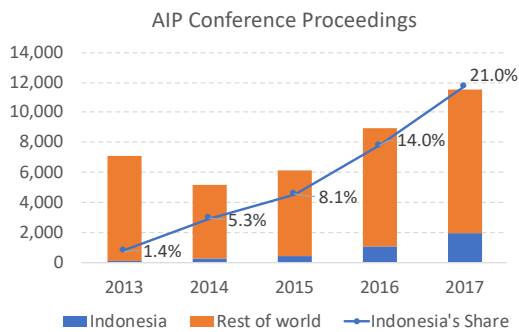

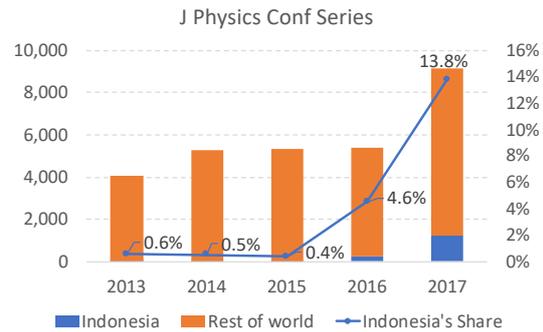

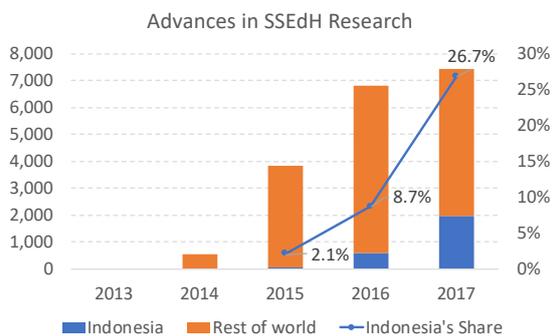

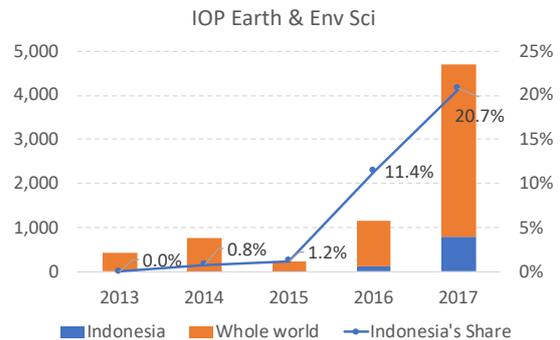

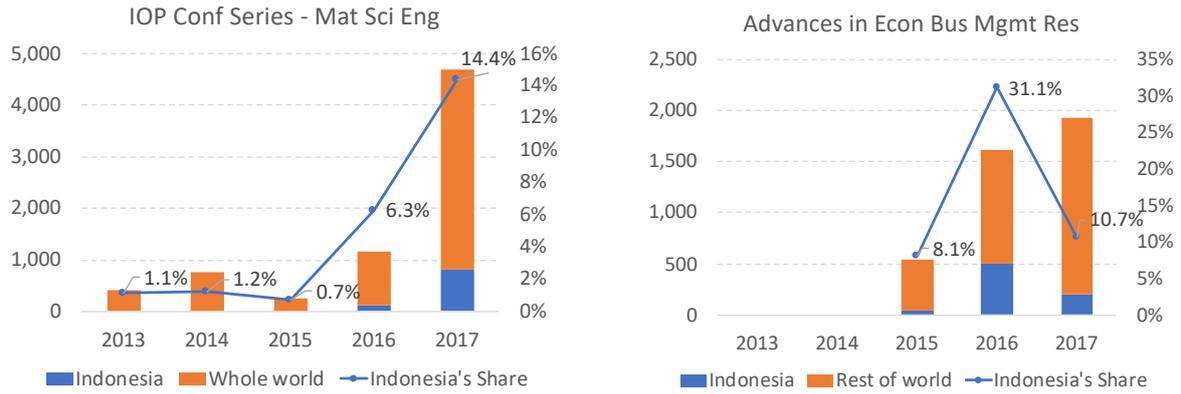

Figure 6.  Indonesian papers in the CPCI by conference

In each case Indonesia's share of the conference proceedings papers has increased in the past five years (Fig. 6). Of the two series with consistent quantity of proceedings papers over the past five years, Indonesia based authors increased output by 20-fold in AIP Conference Proceedings. Conference papers in this series from the rest of the world grew relatively modestly, meaning that by 2017 more than one-fifth (21%) of the papers featured an author from Indonesia. In the Journal of Physics Conference Series, Indonesian authorship growth was even more pronounced and more concentrated in the last two years. Between 2015 and 2017 Indonesia's contribution to this series grew more than 6000% and in 2017 Indonesian authors featured on 13.8% of the conference papers.

Only two of the six conference series studied were added to the WoS after 2013. In one of these (ASSEHR-Advances in Social Sciences, Education & Humanities Research), Indonesian papers were not present in the first year. In the other case (AEBMR-Advances in Economics, Business & Management Research), Indonesian scientists authored fewer than 10% of the resulting proceedings papers in its first year (Fig. 6). In both of those two series, by 2017, Indonesia was ranked second country by author and hosted nearly half (48.8%) and over one-third (36.4%) of the conferences in Indonesia respectively.

Table 4. Indonesia's rank as country publishing most proceedings papers

| Year | AIP Conf Proc Rank | AIP Conf Proc % conferences hosted in Indonesia | J Phys Conf Rank | J Phys Conf % conferences hosted in Indonesia | ASSEHR Rank | ASSEHR % conferences hosted in Indonesia |
|---|---|---|---|---|---|---|
| 2013 | 18 | 3.2 | 31 | 1.4 | | |
| 2014 | 8 | 7.6 | 41 | 2.2 | none | 22.2 |
| 2015 | 5 | 9.8 | 47 | 1.1 | 2 | 7.4 |
| 2016 | 3 | 22.5 | 8 | 6.3 | 2 | 18.4 |
| 2017 | 1 | 20.5 | 2 | 11.9 | 2 | 48.8 |

| Year | IOP EARTH ENV Rank | IOP EARTH ENV % conferences hosted in Indonesia | IOP MAT ENG Rank | IOP MAT ENG % conferences hosted in Indonesia | AEBMR Rank | AEBMR % conferences hosted in Indonesia |
|---|---|---|---|---|---|---|
| 2013 | | | 17 | 7.7 | | |
| 2014 | 19 | 16.7 | 23 | 6.3 | | |
| 2015 | 14 | 16.7 | 27 | 3.0 | 2 | 20.0 |
| 2016 | 4 | 28.6 | 5 | 8.5 | 2 | 25.0 |
| 2017 | 2 | 35.7 | 2 | 14.0 | 2 | 36.4 |

Table 4 shows Indonesia's rank in the list of counties with most authors on the conference papers collected in each series – which are collections of proceedings from multiple conferences in a similar field. Also shown for each series is the proportion of the conferences that were physically hosted in Indonesia. By the year 2017, Indonesia was ranked as first or second country by authorship in all six of the conferences shown. In the case of AIP Conference Proceedings, Indonesia was the highest ranked country by number of author affiliations in 2017, when it hosted more than one-fifth (20.5%) of the conferences in Indonesia, having risen from 18[th] rank on authorship and hosting only 3.2% of the conferences in 2013. IOP's Journal of Physics Conferences showed a comparable pattern with Indonesia appearing as second ranked country by author affiliation by 2017 when 11.9% of the events took place in Indonesia, up from 31[st] authorship position and hosting 1.4% conferences four years earlier. Further details are shown in Table A2 of the Appendix.

In all the six series studied, there was an increasing proportion of the conferences hosted in Indonesia. It is undeniably easier to attend a conference hosted in one's own country because it by-passes the need for international travel which might require visas, authorisations, additional time and expense. Therefore, we could reasonably expect a greater number of attendees, abstract submissions and resulting conference proceedings from Indonesian scientists at conferences held in Indonesia. It therefore follows that the location of the conference might influence the number and proportion of the proceedings papers from a particular country, especially the host country.

The data presented in Table 4 shows Indonesia's rank in the countries with most author affiliations on the proceedings papers in six conferences series by year. This means the publication year, and not necessarily the year the conference took place. The third column shows the proportion of conferences in the series that were hosted in Indonesia. The year the conference was hosted did not always coincide with the year in which the proceedings were published. Sometimes a conference is held at the end of one year, say in December and even a quick publisher will publish the conference proceedings in the following year. That means the two columns in Table 4 are not infallibly related although I have made the assumption that in the majority of cases the relationship holds true. Even so, the data clearly shows an increase in the proportion of conference papers published by Indonesia based authors and an increase in the proportion of conference hosted in Indonesia.

**Indonesia's academic policy and promotion guidelines**

Table 5 shows the journal categories used to determine the 'quality' of publications when academics apply for appointment or promotion to an academic role in Indonesia. These regulations are described in Chapter 5, Article 10 of the Technical instructions for the implementation of academic position by credit numbers (Ministry of Education and Culture 2014). The regulations refer to the level of journal in which a work should be published, but make no reference to document type. Therefore, a conference paper was awarded the same value as a journal article.

Table 5. Journal types considered academic for promotion in Indonesia

| | | Journal quality → | | | |
|---|---|---|---|---|---|
| No. | Academic Position | National journal | Accredited national journal | International Journal | Reputable International Journal |
| 1 | Lecturer | ✓ | | | |
| 2 | Assistant Professor | ✓ | | | |
| 3 | Associate Professor (Master's) | | | ✓ | |
|   | Associate Professor (PhD) | | ✓ | | ⇉ |
| 4 | Professor | | | | ✓ ⇉ |

✓ Initial appointment/Promotion

⇉ Skip position

Notes:
The ticks denote minimum journal categories in which applicants should have published works to be considered for promotion to the relevant academic position. Journal quality is assumed to increase as one moves to the right of the table.
Skip position means jumping two levels above one's current academic level, e.g. from Lecturer to Associate Professor or from Assistant Professor to Professor.
Academics require a higher degree (Master's or PhD) to attain the position of Associate Professor. Associate Professors with a Master's degree have more stringent criteria to be promoted than those who already hold a PhD.

All academics are expected to publish research papers as a core part of their function and in some countries, this is mandated. Indonesian law passed in 2012 (Ministry of Higher Education 2012) encouraged researchers to disseminate results via 'scientific publications' in accredited scientific journals by linking such publications to promotion and awards. Government regulations from 2014 show that academics' eligibility to be appointed or promoted to an academic post are assessed based on a number of credits that can be accumulated by publishing scientific works in journals ranging from national to internationally reputable, and advancement was quicker if more works were published, especially in the higher valued journals (Ministry of Education and Culture 2014). For each academic, the case is first examined by the University, and then reviewed by the Higher Education Office of the Republic of Indonesia.

In both the 2012 and the 2014 regulations, the document type of 'published works' was not defined and there were also incentives for scientific works in 'other forms' and for presenting at conferences so it is reasonable to suppose that academics found avenues for promotion by presenting at a conference and publishing the paper in the associated proceedings more quickly and easily than by publishing full research papers in approved journals. According to guidelines issued in 2017 by the Ministry of Research, Technology and Higher Education, Indonesia has the potential to increase its publication rate through its 250,000 academic researchers at 4,000 higher education institutions each of whom has the legal obligation to conduct and publish scientific studies (Ministry of Research Technology and Higher Education 2017). The 2017 guidelines stated as a national goal increased academic output specifically with respect to Thailand and Malaysia.

Under promotional guidelines released in April 2019, academics will be assessed via a credit system that will determine their entry rank, their promotion rate and even whether they keep their full salary and benefits (Ministry of Research & Higher Education 2019). For the first time, there was a difference in credits awarded between the published document types, for instance a scholar will receive up to 30 credits for conference proceedings if these are indexed in Scopus and Scimago and a maximum of 40 for an article in an impact factor journal.

The 2017 guidelines also referred to an Overseas Conference Attendance Programme that provides support intended to increase published research papers from Indonesian scientists. In addition, the document included precise instructions on how to host an international academic conference in Indonesia and ensure the conference proceedings are indexed in WoS and Scopus. In this section (Chapter 4, p. 66) There was also a list of recommended publishers that a conference organizer could submit their proceedings to. This list includes all those publishers of the six conference series that appeared in this study, and which contained the greatest number of CPCI-indexed proceedings papers with Indonesian authors. The same conference series also showed a steep growth in the number of conferences hosted in Indonesia.

## Conclusions

Recent years have seen a disproportionate increase in conference proceedings publications from Indonesian academics. Neither the world average, regional average nor any peer country observed showed a similar pattern, and that made Indonesia an interesting case study. The growth seen in the WoS was recent, sustained and confirmed by data from Scopus, with as yet inconclusive supporting data from Dimensions. The source of the additional conference papers reflected real growth in published conference papers rather than a result of large conferences or conference series being added to the databases.

The laws, publishing guidelines and credit-based assessment system might therefore have provided the conditions and stimulus for Indonesian academics to increase their publication output and advance their careers partially through a preference for publishing conference papers. Those same policies could have also incentivized scientists to host international conferences in Indonesia which has possibly led to an increase in the proportion of local authors in the proceedings of these conferences.

It does not appear from the policy documents that there is a specific campaign targeting conference proceedings, although the reward system might have permitted scientists to take advantage of this publication practice in pursuit of career advancement. New guidelines released in April 2019 have now distinguished between publication types providing greater reward for journal articles than conference proceedings papers. Future studies will show whether publication behaviour alters accordingly.

## Acknowledgements
Ton van Raan and Ludo Waltman for inspiring me to embarque upon this journey and guiding me expertly through it.
Roy Hendroko Setyobudi for advice on the Indonesian higher education system.

## Competing interests
The author is a Director of Knowledge E, which is an open access publisher of journals and conference proceedings including from Indonesia.

# Appendix: Additional data tables

Table A1. Conference paper output and CAGR by ASEAN countries and regional and global totals

| Country/territory | CAGR | 2008 | 2009 | 2010 | 2011 | 2012 | 2013 | 2014 | 2015 | 2016 | 2017 |
|---|---|---|---|---|---|---|---|---|---|---|---|
| Indonesia | 43.7% | 365 | 479 | 498 | 462 | 936 | 1,484 | 2,228 | 4,189 | 6,289 | 13,735 |
| Brunei | 29.7% | 6 | 12 | 12 | 9 | 24 | 31 | 58 | 75 | 94 | 81 |
| Philippines | 17.3% | 196 | 206 | 205 | 201 | 269 | 297 | 393 | 661 | 606 | 968 |
| Vietnam | 13.7% | 370 | 405 | 323 | 275 | 480 | 697 | 794 | 1,102 | 1,245 | 1,341 |
| Myanmar | 13.4% | 29 | 38 | 23 | 16 | 15 | 8 | 10 | 24 | 72 | 102 |
| **ASEAN** | **13.2%** | **9,573** | **9,861** | **8,682** | **9,177** | **13,126** | **15,070** | **19,382** | **20,998** | **23,233** | **33,029** |
| Malaysia | 12.6% | 2,798 | 3,275 | 2,754 | 3,903 | 5,832 | 6,678 | 8,981 | 7,741 | 7,663 | 9,171 |
| Cambodia | 12.4% | 27 | 27 | 14 | 17 | 17 | 7 | 12 | 29 | 19 | 87 |
| Laos | 5.1% | 20 | 14 | 5 | 8 | 7 | 4 | 18 | 5 | 23 | 33 |
| Thailand | 3.9% | 2,446 | 2,097 | 1,950 | 1,791 | 2,353 | 2,444 | 3,094 | 2,870 | 2,768 | 3,591 |
| Singapore | 2.9% | 3,490 | 3,482 | 3,065 | 2,685 | 3,400 | 3,604 | 4,095 | 4,672 | 4,874 | 4,632 |
| **World** | **2.2%** | **507,711** | **509,176** | **456,521** | **455,251** | **510,427** | **518,939** | **565,863** | **581,959** | **611,135** | **630,291** |

Table A2. Conference proceedings publications, country rank and host location

## AIP Conference Proceedings

| | Conference Proceedings | | | | Host location | | |
|---|---|---|---|---|---|---|---|
| Year | Indonesia | World | Share | Rank | Indonesia | Total | % Indonesia |
| 2006 | 1 | 6,138 | 0.0% | 72 | | | |
| 2007 | 9 | 8,203 | 0.1% | 64 | | | |
| 2008 | 35 | 7,150 | 0.5% | 41 | 1 | 118 | 0.8% |
| 2009 | 26 | 9,013 | 0.3% | 48 | 2 | 115 | 1.7% |
| 2010 | 127 | 9,257 | 1.4% | 22 | 3 | 125 | 2.4% |
| 2011 | 45 | 7,483 | 0.6% | 32 | 1 | 94 | 1.1% |
| 2012 | 141 | 7,873 | 1.8% | 17 | 2 | 90 | 2.2% |
| 2013 | 101 | 7,138 | 1.4% | 18 | 2 | 62 | 3.2% |
| 2014 | 260 | 5,206 | 5.0% | 8 | 5 | 66 | 7.6% |
| 2015 | 461 | 6,124 | 7.5% | 5 | 6 | 61 | 9.8% |
| 2016 | 1,095 | 8,944 | 12.2% | 3 | 20 | 89 | 22.5% |
| 2017 | 1,998 | 11,526 | 17.3% | 1 | 26 | 127 | 20.5% |

## Journal of Physics conference series

| | Conference proceedings | | | | Host location | | |
|---|---|---|---|---|---|---|---|
| Year | Indonesia | World | Share | Rank | Indonesia | Total | % Indonesia |
| 2006 | | 1,839 | | | | | |
| 2007 | | 1,502 | | | | | |
| 2008 | | 3,059 | | | | | |
| 2009 | | 3,223 | | | | | |
| 2010 | 6 | 4,353 | 0.1% | 53 | 0 | 62 | 0.0% |
| 2011 | 1 | 4,217 | 0.0% | 70 | 0 | 76 | 0.0% |

| Year | Indonesia | World | Share | Rank | Indonesia | Total | % Indonesia |
|---|---|---|---|---|---|---|---|
| 2012 | 1 | 5,079 | 0.0% | 83 | 0 | 71 | 0.0% |
| 2013 | 24 | 4,071 | 0.6% | 31 | 1 | 72 | 1.4% |
| 2014 | 28 | 5,285 | 0.5% | 41 | 2 | 93 | 2.2% |
| 2015 | 21 | 5,342 | 0.4% | 47 | 1 | 92 | 1.1% |
| 2016 | 246 | 5,406 | 4.6% | 8 | 7 | 112 | 6.3% |
| 2017 | 1,265 | 9,172 | 13.8% | 2 | 20 | 168 | 11.9% |

**Advances in social science education and humanities research**

| | Conference proceedings | | | | Host location | | |
|---|---|---|---|---|---|---|---|
| Year | Indonesia | World | Share | Rank | Indonesia | Total | % Indonesia |
| 2006 | | | | | | | |
| 2007 | | | | | | | |
| 2008 | | | | | | | |
| 2009 | | | | | | | |
| 2010 | | | | | | | |
| 2011 | | | | | | | |
| 2012 | | | | | | | |
| 2013 | | | | | | | |
| 2014 | | 544 | | | 2 | 9 | 22.2% |
| 2015 | 81 | 3831 | 2.1% | 2 | 2 | 27 | 7.4% |
| 2016 | 593 | 6794 | 8.7% | 2 | 7 | 38 | 18.4% |
| 2017 | 1992 | 7452 | 26.7% | 2 | 42 | 86 | 48.8% |

**IOP Conference Series - Earth and Environmental Science**

| | Conference Proceedings | | | | Host location | | |
|---|---|---|---|---|---|---|---|
| Year | Indonesia | World | Share | Rank | Indo | Total | % Indonesia |
| 2006 | | | | | | | |
| 2007 | | | | | | | |
| 2008 | | 67 | | | | | |
| 2009 | | 43 | | | | | |
| 2010 | | 201 | | | | | |
| 2011 | | 11 | | | | | |
| 2012 | | | | | | | |
| 2013 | | 425 | | | | | |
| 2014 | 6 | 766 | 0.8% | 19 | 1 | 6 | 16.7% |
| 2015 | 3 | 242 | 1.2% | 14 | 1 | 6 | 16.7% |
| 2016 | 118 | 1,038 | 11.4% | 4 | 6 | 21 | 28.6% |
| 2017 | 806 | 3,885 | 20.7% | 2 | 20 | 56 | 35.7% |

**IOP Conference series - Materials Science and Engineering**

| | Conference Proceedings | Host location |
|---|---|---|

| Year | Indonesia | World | Share | Rank | Indo | Tot | % Indonesia |
|---|---|---|---|---|---|---|---|
| 2006 | | | | | | | |
| 2007 | | | | | | | |
| 2008 | | 35 | 0.0% | | | | |
| 2009 | | 173 | 0.0% | | | | |
| 2010 | | 611 | 0.0% | | | | |
| 2011 | 1 | 418 | 0.2% | 33 | 0 | 11 | 0.0% |
| 2012 | 1 | 618 | 0.2% | 59 | 0 | 15 | 0.0% |
| 2013 | 7 | 626 | 1.1% | 17 | 1 | 13 | 7.7% |
| 2014 | 8 | 655 | 1.2% | 23 | 1 | 16 | 6.3% |
| 2015 | 14 | 2046 | 0.7% | 27 | 1 | 33 | 3.0% |
| 2016 | 193 | 3080 | 6.3% | 5 | 5 | 59 | 8.5% |
| 2017 | 1179 | 8215 | 14.4% | 2 | 19 | 136 | 14.0% |

**Advances in Economics Business and Management Research**

| | Conference Proceedings | | | | Host location | | |
|---|---|---|---|---|---|---|---|
| Year | Indonesia | World | Share | Rank | Indo | Total | % Indonesia |
| 2006 | | | | | | | |
| 2007 | | | | | | | |
| 2008 | | | | | | | |
| 2009 | | | | | | | |
| 2010 | | | | | | | |
| 2011 | | | | | | | |
| 2012 | | | | | | | |
| 2013 | | | | | | | |
| 2014 | | | | | | | |
| 2015 | 44 | 545 | 8.1% | 2 | 1 | 5 | 20.0% |
| 2016 | 503 | 1617 | 31.1% | 2 | 4 | 16 | 25.0% |
| 2017 | 205 | 1924 | 10.7% | 2 | 8 | 22 | 36.4% |